\documentclass[pra,aps,twocolumn,groupedaddress,superscriptaddress]{revtex4-2}
\usepackage[usenames,dvipsnames]{color}
\usepackage[colorlinks=true,linkcolor=Blue,citecolor=Blue,urlcolor=Blue]{hyperref}
\usepackage{graphicx}
\usepackage{bbm}
\usepackage{bm}
\usepackage{amsmath}
\usepackage{amssymb}
\usepackage{mathptmx}
\usepackage[version=3]{mhchem}

\newcommand{\ba}{\begin{eqnarray}}

\newcommand{\ea}{\end{eqnarray}}
\newcommand{\bd}{\begin{displaymath}}

\newcommand{\nn}{\nonumber \\}

\newcommand{\bs}{\bigskip}

\begin{document}
\title{Fractonic Quantum Quench in Dipole-constrained Bosons}

\author{Yun-Tak \surname{Oh}}
\affiliation{Division of Display and Semiconductor Physics, Korea University, Sejong 30019, Korea}

\author{Jung Hoon \surname{Han}}
\affiliation{Department of Physics, Sungkyunkwan University, Suwon 16419, South Korea}

\author{Hyun-Yong \surname{Lee}}
\email{hyunyong@korea.ac.kr}
\affiliation{Division of Display and Semiconductor Physics, Korea University, Sejong 30019, Korea}
\affiliation{Department of Applied Physics, Graduate School, Korea University, Sejong 30019, Korea}
\affiliation{Interdisciplinary Program in E$\cdot$ICT-Culture-Sports Convergence, Korea University, Sejong 30019, Korea}

\begin{abstract} 
We investigate the quench dynamics in the dipolar Bose-Hubbard model (DBHM) in one dimension. The boson hopping is constrained by dipole conservation and show fractonic dynamics. The ground states at large Hubbard interaction $U$ are Mott insulators at integer filling and a period-2 charge density wave (CDW) at half-integer filling. We focus on Mott-to-Mott and CDW-to-CDW quenches and find that dipole correlation spreading shows the light-cone behavior with the Lieb-Robinson (LR) velocity proportional to the dipole kinetic energy $J$ and the square of the density in the case of Mott quench at integer filling. Effective model for post-quench dynamics is constructed under the dilute-dipole approximation and fits the numerical results well. For CDW quench we observe a much reduced LR velocity of order $J^2/U$ and additional periodic features in the time direction. The emergence of CDW ground state and the reduced LR velocity at half-integer filling can both be understood by careful application of the second-order perturbation theory. The oscillatory behavior arises from quantum scars in the quadrupole sector of the spectrum and is captured by a PXP-like model that we derive by projecting the DBHM to the quadrupolar sector of the Hilbert space. 
\end{abstract}

\date{\today}
\maketitle

{\it Introduction.}- Dipole-conserving systems are a simple example of the particle dynamics and the many-body phases being altered in a fundamental way by kinetic constraints\,\cite{sachdev02,prem18,pollmann19,refael19,pollmann20,sala2020ergodicity,feldmeier20,nandkishore20,gromov20,nandkishore21,moudgalya20prb,moudgalya2022,gorantla22,moudgalya22prx,feldmeier23,lake1,lake23,skinner23,bakr20,aidelsburger21,aidelsburger23,weitenberg22,ageev2023-1,feldmeier-quench}. In addition to the immobility of single particles reminiscent of the fractonic dynamics, other novel phenomena such as the lack of thermalization, Hilbert space fragmentation, and quantum scars are all manifested in the dipole-constrained systems\,\cite{pollmann19,refael19,papic21,moudgalya20prb,moudgalya2022,skinner23,aidelsburger23,pan23}. More recently they have received a great deal of attention as ways to understand anomalous transport and relaxation phenomena in tilted optical lattices\,\cite{bakr20,aidelsburger21,weitenberg22,aidelsburger23,pan23}. 

Over the years the optical lattice system has proven to be excellent platforms for probing non-equilibrium states of matter. A prototypical example of non-equilibrium probe is the quench dynamics where a sudden change of system parameters results in the ground state evolving according to the post-quench Hamiltonian. Some intriguing aspects of the post-quench dynamics have been examined in the past, ranging from light cone-like information spreading subject to the Lieb-Robinson bounds\,\cite{lauchli07,lauchli08,bloch12,kollath12,saito21,lucas22}, dynamical quantum phase transition (DQPT)\,\cite{heyl13,DQCP-in-ions17}, and quantum scars\,\cite{bernien17,turner18,serbyn21,papic21}. The issues have been addressed in the framework of e.g. Bose-Hubbard model\,\cite{lauchli07,lauchli08,kollath12}, transverse Ising model\,\cite{DQCP-in-ions17}, and PXP model\,\cite{turner18, papic21}. 

Motivated by recent experiments in tilted optical lattices, several interacting models embodying the dipole conservation in addition to the charge conservation have been proposed\,\cite{aidelsburger21,lake1,lake23,feldmeier23,feldmeier-quench}. An interesting ramification of one such model, called the dipolar Bose-Hubbard model (DBHM)\,\cite{lake1,lake23,feldmeier23}, is the disappearance of conventional superfluid phase and the emergence of dipole condensate phase taking its place in the weak Hubbard interaction regime. The ground state phase diagrams and various low-energy correlations of this model have been worked out. Notably, single-particle correlations are heavily suppressed in all phases of the model and two-particle dipole-dipole correlations take over as a measure of (quasi-)ordering. Recent progress in experiments shows that DBHM and its fermionic cousin, the dipolar Fermi-Hubbard model, are among the most experimentally accessible models displaying fractonic quasiparticle behavior through enforcing the dipole symmetry\,\cite{bakr20,aidelsburger21,weitenberg22,aidelsburger23,pan23}. 

Despite the growing importance of dipole-constrained models with roots in tilted optical lattice, the quench dynamics of DBHM has not been examined theoretically. Here we present the first thorough study of the quench dynamics over different phases of DBHM at integer and half-integer fillings. Due to the strict prohibition of single-particle dynamics, dipoles as low-energy excitations become the main channel of correlation spreading. The Lieb-Robinson (LR) bound for the Bose-Hubbard model, which scales linearly with the density, is replaced by a new bound scaling as the {\it square of the density} in DBHM. At half-integer filling where the ground state is a period-2 charge-density wave (CDW), dipole correlation spreading is bounded by a much smaller LR speed and a periodic (in time) revival, reminiscent of quantum scars. Effective models for the post-quench dynamics in both integer and half-filling fillings are derived in terms of a low density of dipole excitations and can explain the numerically observed LR bound quantitatively. Furthermore, a PXP-like model consistent with the scar-like features in the half-integer quench can be derived by taking into account quadrupole excitations, and explain the observed periodicity very well. 
\\


{\it Model and methods.}- The one-dimensional DBHM is\,\cite{lake1,feldmeier23,lake23}
\begin{align}
    H = -J \sum_x ( b^\dag_{i-1} b_x^2 b^\dag_{x+1} + h.c.)  + \frac{U}{2} \sum_x n_x (n_x -1) , \label{eq:boson-pair-hop} 
\end{align}
where $n_x = b^\dag_x b_x$ is the boson number at site $x$, and $n = \sum_x n_x /L$ ($L$=number of sites) is the average density. The key departure from BHM is the absence of one-boson hopping and the dipolar hopping ($J$) that takes its place. The model is invariant under both the global U(1) and the dipolar U(1) phase changes $b_x \rightarrow e^{i\theta}b_x , ~~    b_x \rightarrow e^{i \theta x } b_x$, and possesses two conserved quantities:  the total charge $Q = \sum_x b^\dag_x b_x$ and the dipole moment $D = \sum_x x b^\dag_x b_x$. 

We employ the density matrix renormalization group\,(DMRG)\,\cite{White1992, White1993, Schollwock2005, Schollwock2011} and the time-dependent variational principle\,(TDVP)\,\cite{Haegeman2011, Haegeman2016} calculations to explore the ground state and its quench dynamics. For DMRG simulations, we utilize the two-site and subspace expansion algorithms\,\cite{Hubig2015}, focusing on a finite system with size $L=100$ and limiting the local boson number at each site to 10. The maximum bond dimension for DMRG is set to $\chi_{\rm DMRG} = 500$ ensuring an accurate representation of the ground state in the matrix product states representation. In the context of TDVP, we adopt both one-site and two-site algorithms, with the maximum bond dimension up to $\chi_{\rm TDVP} = 3000$. This substantial increase in the maximum bond dimension allows for a more detailed exploration of the system's dynamics. We also incorporate the conservation of boson number\,$Q$ and dipole moment\,$D$ in both DMRG and TDVP simulations. It not only guarantees the conservation of associated $U(1)$ symmetries but also greatly enhances the computational efficiency of the simulations\,\cite{Singh2011}.
\\

{\it Mott quench.} --
We obtain the ground state $|\psi\rangle$ of the DBHM\,[Eq.\,\eqref{eq:boson-pair-hop}] at $U=U_i$ and observe their evolution under the new Hamiltonian with $U=U_f$ as $|\psi(t)\rangle = e^{-i tH_{\rm DBHM}} |\psi\rangle$. The final value of $U_f$ is chosen such that the equilibrium state corresponding to $U=U_f$ is also in a Mott phase. We refer to such quench as the Mott-to-Mott quench, or simply Mott quench. In the simulation $J$ is set to unity. With $|\psi (t)\rangle$ we examine the time evolution of relevant quantities such as correlators and fidelities. The single-boson correlation $\langle \psi(t) | b^\dag_x b_{x'} |\psi (t)\rangle$ remains strictly zero except $x=x'$ at all times due to the dipole constraint, indicating the fractonic nature of the single-boson particle in the DBHM. Instead, meaningful information is contained in the {\it dipole correlator}
%
$ C_d (x,t) = {\rm Re} [\langle \psi (t) |d^\dag_{x_0+x} d_{x_0} |\psi(t)\rangle ]$
%
where $d_x = b^\dag_x b_{x+1}$ is the dipole operator. In the TDVP simulation we choose $x_0 = L/2$ at the center of the system $1 \le x \le L$; results are unaffected by the choice of $x_0$ unless it is positioned too close to the boundary.

\begin{figure}
    \centering
    \includegraphics[width=0.5\textwidth]{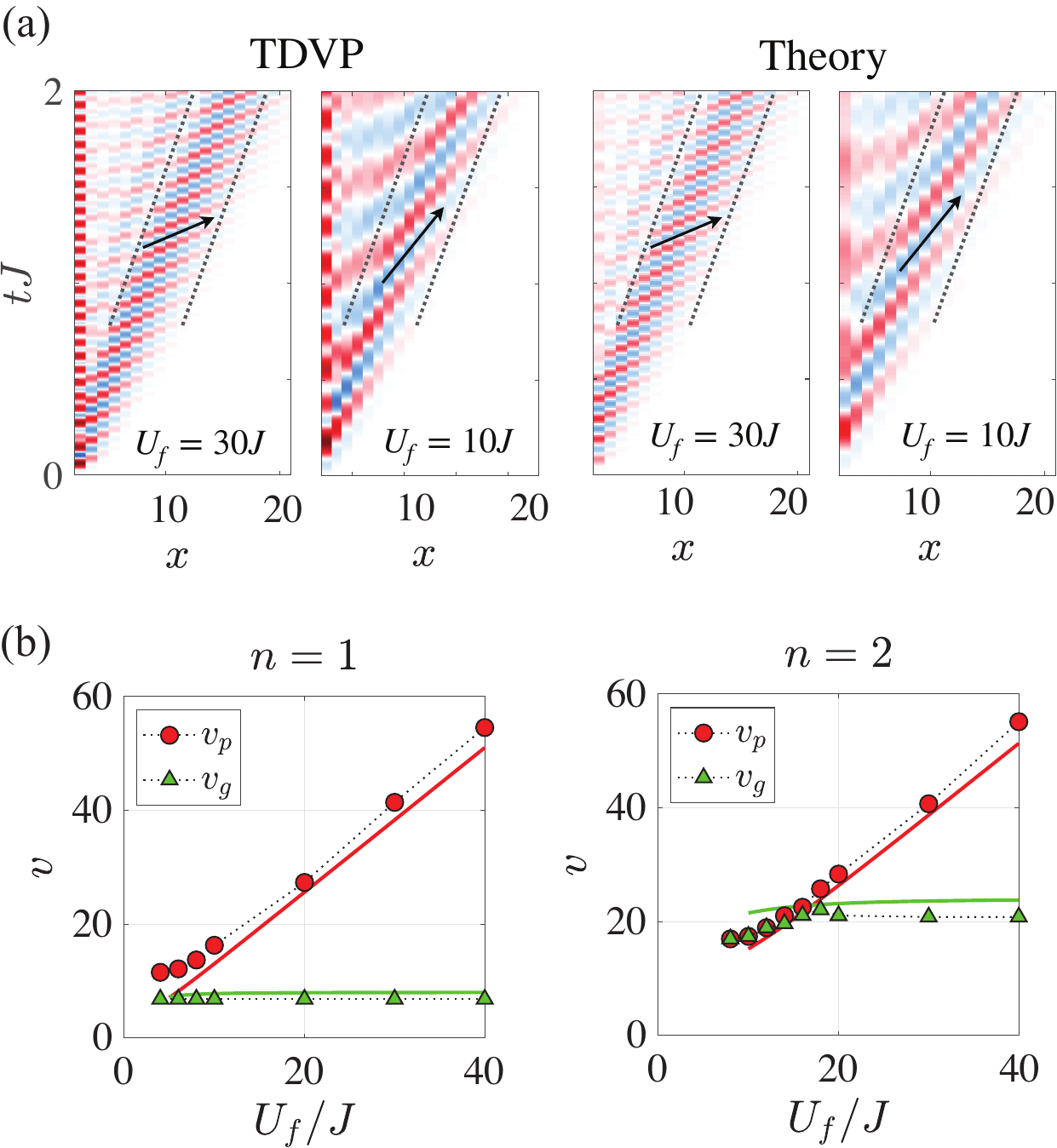}
    \caption{
    (a) TDVP and effective model results of the post-quench dipole correlation function (real part) $C_d(x,t)$ at the filling $n=1$ and $2$ with the initial value $U_i /J = 100$. The data is normalized such that the maximum value is adjusted to unity.
    The dashed line and the solid arrows represent the travel speed of the overall wave packet and the peak in the response, respectively. (b) Phase and group velocities for the dipole correlation at $n=1$ and $2$ as a function of the quench interaction $U_f/J$ deduced from the TDVP data such as shown in (a). 
    }
    \label{fig:mott_quench}
\end{figure}

We begin by focusing on the integer-filling $n_x = n$ at large $U/J$ where the ground state is a Mott state, faithfully  represented as a product state $|{\rm M}\rangle \equiv \otimes_{x=1}^L |n\rangle_x $. The dipole correlation function in the Mott state is extremely short-ranged, but 
the quench triggers the spreading of the correlation with a well-defined propagation front in the shape of a light-cone as shown in Fig.\,\ref{fig:mott_quench}\,(a). A modern interpretation of this is in terms of the LR bound\,\cite{lieb72}, whose existence has been rigorously proven for the conventional BHM\,\cite{saito21,lucas22,faupin22,lucas23} after many years of numerical observation to the effect\,\cite{lauchli08,bloch12,kollath12,langen13,heyl19,danshita22}. The light-cone spreading of dipole correlation in Fig.\,\ref{fig:mott_quench}\,(a) is highly suggestive of an LR bound in the DBHM as well, with the information carried in the dipole, not charge, sector. To make a quantitative statement on the LR bound of the DBHM, we extract the group velocity\,($v_g$) and also the phase velocity\,($v_p$) from the TDVP data by fitting the leading wave packets in the dipole correlation\,\cite{Malcolm2010}. See Supplementary Material (SM) for details on how to determine the velocities\,\cite{sm}. The results are presented in Fig.\,\ref{fig:mott_quench}\,(b) as a function of $U_f$ at filling $n=1$ and $2$, strongly indicating that $v_g$ remains independent of $U_f$, whereas $v_p$ exhibits a linear dependence on it. In terms of the filling factor dependence, $v_g$ appears to increase with the square of the filling factor, while $v_p$ remains independent of it. 

The Mott quench dynamics can be comprehensively understood by developing an effective model deep inside the Mott phase $U\gg J$. The low-lying excitations in the Mott phase are the two kinds of dipole excitations $|l_x\rangle \sim d_x |{\rm M}\rangle$ and $|r_x\rangle \sim d_x^\dagger |{\rm M}\rangle$ called $l$-dipoles and $r$-dipoles, respectively. Considering a Hilbert subspace consisting of the Mott state $|{\rm M}\rangle$ and the dipoles $\{ |l_x \rangle, | r_y\rangle \}$, the effective Hamiltonian in this space can be derived\,\cite{sm} 
\begin{align}
    H_{\rm eff} =&  \sum_{k,\sigma} \omega_k \gamma_{k\sigma}^\dagger \gamma_{k\sigma} , & 
    \omega_k = & (\rho_k^2  + | \lambda_k |^2)^{1/2}.
    \label{eq:dipole-hamil-omega}
\end{align}
The parameters $\rho_k$ and $\lambda_k$ are given by\,\cite{sm}
\begin{align}
\rho_k & = U - 2 J n (n+1) \cos k, \nn
\lambda_k e^{i \mu_k } & = J \left( n \sqrt{(n+1)(n+2)} e^{ik} - (n+1) \sqrt{n (n-1)} e^{-ik} \right)
\nonumber
\end{align}
for general integer filling factor $n$. The operator $\gamma^\dag_{k\sigma}$ creates Bogoliubov quasi-particles with pseudo-spin $\sigma=l,r$ and momentum $k$. There is much resemblance of this effective model to the quasiparticle model for Mott quench in the BHM\,\cite{bloch12,kollath12}, with the key difference that dipoles rather than doublons and holons are the elementary excitations here. 

The post-quench wavefunction $|\psi(t)\rangle$ can be derived in exact form using the Bogoliubov Hamiltonian and  allows a closed-form expression of the dipole correlation function at $t>0$\,\cite{sm}, which shows excellent agreement with the TDVP simulations as shown in Fig.\,\ref{fig:mott_quench}\,(a). 
One can deduce the two propagation velocities analytically from the Bogoliubov model as follows\,\cite{sm}: 
\begin{align}
    & v_g \equiv \max_{k = k_{\rm max}} \left(2 \partial_k \omega_k\right) = 4Jn(n+1) + O\left(\frac{J^3}{U^2}\right), \nonumber\\
    & v_p \equiv  \frac{2 \omega_{k_{\rm max}} }{ k_{\rm max} } = \frac{4U}{\pi}  + O\left(\frac{J^3}{U^2} \right).
    \label{eq:velocities}
\end{align}
When $U\gg J n^2 $, $v_g$ is maximized at $k_{\rm max} =  \pi/2$. The two velocity expressions provide very good fits to the velocities extracted from the TDVP data, as shown in Fig.\,\ref{fig:mott_quench}\,(b). Furthermore, the quadratic dependence of the group velocity on the density $v_g \sim n(n+1)$ deduced from effective model captures the observed increase in $v_g$ by three times in going from $n=1$ to $n=2$. This contrasts with its linear dependence on $n$ in the conventional BHM\,\cite{kollath12, lucas22}. The group velocity in the DBHM approaches zero as $n\rightarrow 0$, whereas it remains finite in the conventional BHM\,\cite{lucas22,lucas23}.
\\
\begin{figure}
    \centering
    \includegraphics[width=0.5\textwidth]{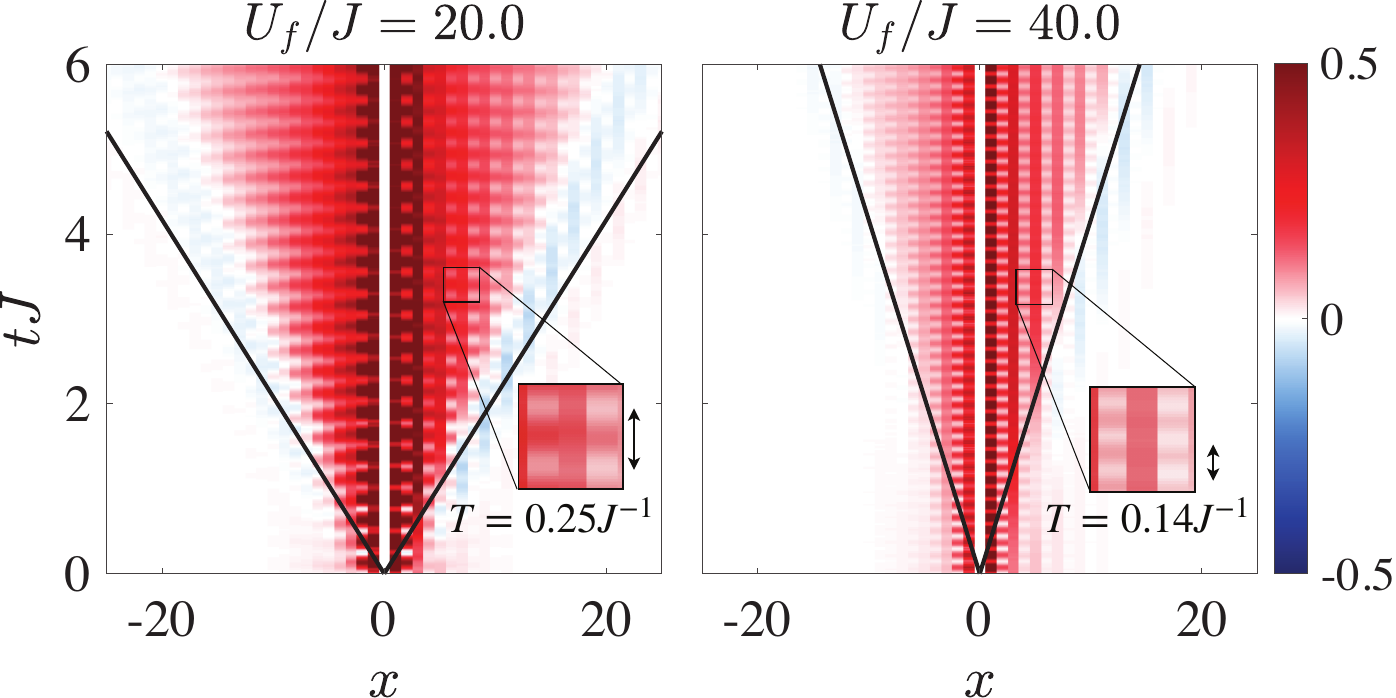}
    \caption{
    TDVP results of the dipole correlation $C_d(x,t)$ at the filling $n=3/2$, with the initial $U_i/J = 100$. The Data is normalized such that the maximum value is adjusted to unity. The black solid line indicates the wavefront expected from the theory\,(See text).
    }
    \label{fig:cdw_quench}
\end{figure}

{\it CDW quench.} -- The correlation spreading at half-integer filling $n+1/2$ shows a number of features which distinguish it sharply from those in the Mott quench. Though the discussion is based on detailed numerics at $n=3/2$, the results straightforwardly generalize to arbitrary half-integer filling. Firstly, the ground state at half-filling obtained by DMRG is a period-2 CDW state with an alternate occupation of one boson and two bosons per site. The LR velocity bounding the correlation spreading in the CDW quench scales as $J^2/U_f$ and substantially smaller than the Mott-quench value which scales as $J$. Finally, the dipole correlation functions show a periodic revival in time that was absent in the Mott quench. Both these features are apparent in the plots shown in Fig.\,\ref{fig:cdw_quench}. 

First we discuss the origin of the CDW ground state at half-filling. The Hubbard term at half-filling demonstrates extensive degeneracy, with any state with half the sites occupied by one boson and the other half with two bosons sharing the same Hubbard energy. The massive degeneracy is lifted at the second-order of $J/U$ in degenerate perturbation theory, resulting in two-fold degenerate CDW ground states\,\cite{sm}. Without loss of generality, we choose the CDW state on an open chain of length $L$ to be $|{\rm CDW}\rangle \equiv \otimes_{a=1}^{L/2} |1_{2a-1}, 2_{2a}\rangle$.  

As in the Mott quench, low-lying excitations are those of $l$- and $r$-dipoles, created in equal numbers to preserve the total dipole moment. Due to the translation symmetry breaking of the CDW, however, the $l$-dipoles ($r$-dipoles) are created at odd (even) sites only, given by the change in the local occupation 
\begin{align} 
|l_{2a-1} \rangle & \sim d_{2a-1} |1_{2a-1},2_{2a}\rangle \sim |2_{2a-1},1_{2a}\rangle \nn 
|r_{2a} \rangle & \sim d_{2a}^\dag |2_{2a},1_{2a+1}\rangle \sim |1_{2a},2_{2a+1}\rangle. 
\nonumber
\end{align} 
Degenerate perturbation theory leads to an effective Hamiltonian of the dipoles in the CDW state\,\cite{sm}:
\begin{align}
    H^{D}_{\rm eff} = & - 12\frac{ J^2}{U} \sum_{a} \left(\, |l_{2a-1}\rangle \langle l_{2a+1}| + |r_{2a}\rangle \langle r_{2a+2}| + h.c. \right) \nn 
    &+32.8\frac{J^2}{U} \sum_a \left(|l_{2a-1}\rangle \langle l_{2a-1}| + |r_{2a} \rangle \langle r_{2a} | \right) . 
    \label{eq:cdw-eff-h}
\end{align}
The superscript $D$ is a reminder that only the dipole states comprise the low-energy Hilbert space, of order $J^2/U$ above the CDW ground state per dipole, used to construct the effective Hamiltonian. 

In constructing the effective model we ruled out configurations where the $(l,r)$ dipoles are adjacent, i.e. $| r_{2a-2} l_{2a-1}  \rangle \equiv |131_{2a-1} \rangle$ and $|l_{2a} r_{2a+1} \rangle \equiv |202_{2a} \rangle$. They are in fact quadrupole and anti-quadrupole excitations, and cost an energy of order $U$ more than two separately created dipoles. Ignoring the quadrupole events, the effective Hamiltonian \eqref{eq:cdw-eff-h} can be diagonalized with the dispersion $\omega_k = (J^2/U)(32.8 - 24 \cos 2 k ) \ge 8.8 J^2/U$. The factor 2 in $\cos 2k$ appears as a result of the unit cell doubling. The group velocity is deduced $ v_g  = \max ( 2 \partial_{k} \omega_k ) = 96 J^2/U$. This prediction, shown as black solid lines in Fig.\,\ref{fig:cdw_quench}, agrees very well with the TDVP results for the propagation boundary of the dipole correlation function. Being of order $J^2/U$, the LR velocity is considerably smaller than the $v_{\rm LR} \sim J$ in the Mott quench and, moreover, depends inversely on $U$ in marked contrast to the Mott quench at integer filling or the quench in the conventional Bose-Hubbard model where it is governed exclusively by kinetic energy $J$. 

\begin{figure}
    \centering
    \includegraphics[width=0.48\textwidth]{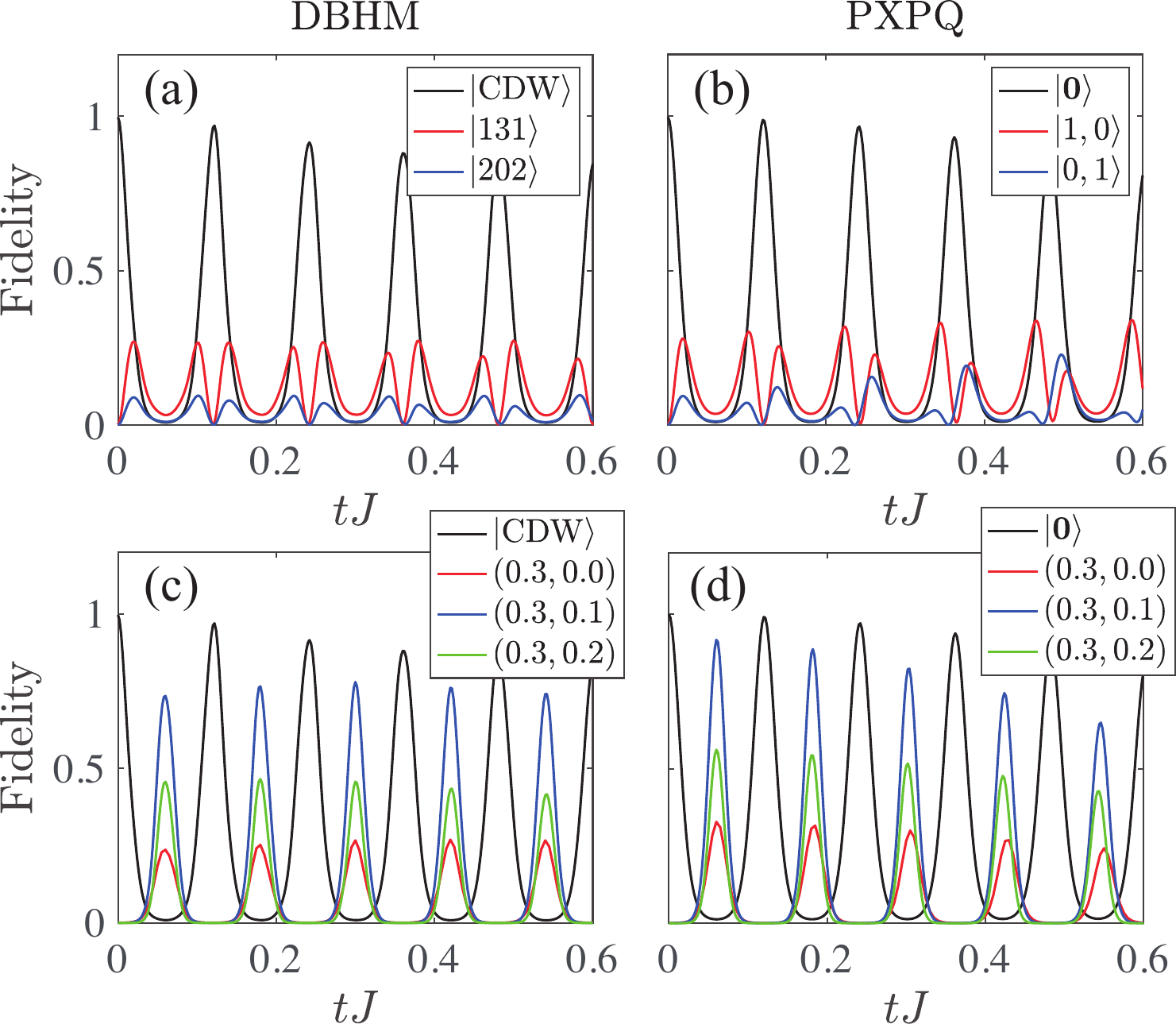}
    \caption{ Dynamics of the quantum fidelity are illustrated in (a) for the DBHM and (b) for the PXPQ model provided in Eq.\,\eqref{eq:cdw-pxp}, where $U_f = 50J$ and $L=200$. In the DBHM, the projections of the post-quench wavefunction $|\psi(t)\rangle$ onto the states $|{\rm CDW}\rangle$, $|131\rangle$, and $|202\rangle$ are displayed, while states $|0\rangle$, $|1,\bar{0}\rangle$, and $|0,\bar{1}\rangle$ are considered in the latter case. Additionally, the projection onto the quadrupole condensate states\,[Eq.\,\eqref{eq:resonating-131}] with fugacities $(\alpha,\beta)$ in the legends are presented in (c) and (d).
    }
    \label{fig:fidelity}
\end{figure}

The other prominent features in the CDW quench, i.e. a periodic revival of the correlation, originates from quadrupole excitations that were previously ignored in deriving the effective model \eqref{eq:cdw-eff-h} in the  dipole subspace. To prove the importance of quadrupole excitations, we calculate the time evolution of fidelities, $F(t) = |\langle \phi |\psi(t)\rangle|^2$ shown in Fig.\,\ref{fig:fidelity}\,(a), measuring the overlap to three target states $|\phi\rangle$: $|{\rm CDW}\rangle$, and the two quadrupole states $|131\rangle = (2/L)^{1/2}\sum_{a}|131_{2a - 1}\rangle$, and $|202\rangle=(2/L)^{1/2}\sum_{a}|202_{2a}\rangle$. All three fidelities undergo periodic revivals with the same frequency as in the dipole correlation function - a phenomenon often observed in scarred systems\,\cite{turner18,serbyn21,yuan22}.

Motivated by the fidelity results, we consider a new subspace in which only the quadrupole states are kept along with the CDW ground state. The connection between dipole and quadrupole sectors are considered negligible, given that such connections are made by an intermediate state of energy $2U$ or more, and are suppressed by a factor of $J/U$ in perturbation theory. On the other hand, quadrupole states are linked to the CDW by a single application of the dipolar kinetic term in DBHM as: 
\begin{align} 
& q_{2a - 2}|{\rm CDW}\rangle = 2\sqrt{6}|131_{2a-1}\rangle, ~
q_{2a - 1}^\dag|{\rm CDW}\rangle = 2 \sqrt{2}|202_{2a}\rangle, 
\nn
& q_{2a - 2}^\dag|131_{2a-1}\rangle = 2 \sqrt{6}|{\rm CDW}\rangle, ~ 
q_{2a - 1}|202_{2a}\rangle = 2 \sqrt{2}|{\rm CDW}\rangle , 
\nonumber
\end{align}
where we introduce the quadrupole operator $q_x = d_x^\dag d_{x+1}$.

The algebra suggests that at every odd site $x = 2a - 1$, 
$\{ |{\rm CDW}\rangle, |131\rangle \}$ forms a two-level system and at every even site $x = 2a$, $\{ |{\rm CDW}\rangle, |202\rangle \}$ forms  another two-level system. 
This structure can be effectively modeled by assigning pseudo-spin-1/2 operators $(X , Z)$ acting on an effective qubit to every site in the lattice: 
\begin{align} 
X_x |n_x \rangle = |n_x + 1 ~ ({\rm mod} ~ 2 ) \rangle, ~~ Z_x |n_x \rangle = (1-2 n_x ) |n_x \rangle . \nonumber
\end{align} 
The CDW state maps to $|{\bf 0}\rangle = \otimes_{x=1}^{L} |0\rangle_x$. Projecting the DBHM to the Hilbert space of quadrupoles gives 
%
\begin{align}
H_J  \rightarrow &  -2J\Bigg(\sqrt{6} \sum_{x \in  {\rm odd} }  P_{x-1} X_x P_{x+1}  \nn &~~~~~~~~~~~~~~~~ + \sqrt{2} \sum_{x \in  {\rm even} } P_{x-2} P_{x-1} X_x P_{x+1} P_{x+2}\Bigg) \nn
H_U  \rightarrow & U\sum_{x=1}^{L}  n_x 
\label{eq:cdw-pxp}
\end{align}
with CDW as the vacuum. The projector $P_x = (1+Z_x)/2$ projects a local state to $|0\rangle_x$. The $P_{x-1} X_x P_{x+1}$ in the first line means that if one tries to create a $131$-quadrupole at the odd site $x$, one can only do so if both of the adjacent sites $(x-1, x+1)$ are devoid of existing $202$-quadrupoles. Otherwise, creating a 131-quadrupole on top of an existing $202$-quadrupole at the adjacent site annihilates the state altogether. 
In the second line, the $X$-operator tries to create a $202$-quadrupole at the even site $x$, provided that its two adjacent sites are devoid of any existing $131$-quadrupoles (hence $P_{x-1}X_x P_{x+1}$). 
Extra projectors $P_{x-2}, P_{x+2}$ are used because creating two $202$-quadrupoles at adjacent positions like $x$ and $x+2$ leads to an occupation of $20402$ and an additional energy cost of $4U$ compared to separate $202$-quadrupoles.
In the first line, the projectors at second-neighbor sites are omitted since generating two $131$-quadrupoles at sites $x$ and $x+2$ yields an occupation of $13031$ without incurring additional energy compared to separate $131$-quadrupoles.
We refer to the emergent projected Hamiltonian as the PXPQ model, with ``Q" referring to the quadrupole excitations. The on-site energy cost $U$ acts as an effective magnetic field polarizing the state toward the CDW\,\cite{pan23,daniel23prb}. Translational symmetry is explicitly broken by two lattice spacings in the PXPQ model.

We calculate the post-quench wave function $|\psi(t)\rangle = e^{-i\,t\,H_{\rm PXPQ} } |{\bf 0}\rangle$ and its overlap with $|1,0\rangle = (2/L)^{1/2} \sum_{x \in {\rm odd}} |1\rangle_x$ and $|0,1\rangle = (2/L)^{1/2} \sum_{x \in {\rm even}} | 1\rangle_x$ using the PXPQ model. For the same values of $(J,U)$, we find very good agreement in the fidelity evolution $F(t)$ as shown in Fig.\,\ref{fig:cdw_quench}\,(b) with the correct period as found in the DBHM. 
Having identified a PXP-like effective model governing the dynamics of quadrupoles, we address the important question of the nature of the quantum scar state in the PXPQ model, by constructing a coherent state of quadrupoles or a quadrupole-condensate (QC)
\begin{widetext}
\begin{align} 
    |{\rm QC} \rangle 
    = & \Omega \, P \, \prod_{x \in {\rm odd}} \left( 1 + \alpha   \, X_x \right) \prod_{x \in {\rm even}} \left( 1 + \beta \, X_x \right)|{\bf 0}\rangle \quad \quad \quad \quad \quad \quad \quad {\rm (in\,\,PXPQ)} \nn 
    = &  \Omega \, P \, \prod_{x\in {\rm odd}} \left( 1 + \frac{\alpha}{2\sqrt{6}}\,\, q_x^\dagger \right) \prod_{x\in {\rm even}} \left( 1 + \frac{\beta}{2\sqrt{2}}\,\, q_x^\dagger \right) |{\rm CDW}\rangle \quad {\rm (in\,\, DBHM)} 
    \label{eq:resonating-131}
\end{align}
\end{widetext}
where $\Omega$ is the normalization, and $\alpha\,(\beta)$ is the fugacity parameter of the 131\,(202) excitation. Here, a projector $P$ rules out configurations that contain overlapping two 202 excitations or neighboring 131-202 excitations. The $|{\rm QC}\rangle$ state can be represented by a matrix product state with a small bond dimension $\chi = 3$ and $4$ in PXPQ and DBHM, respectively\,\cite{sm}. Note that the amplitude $| \langle {\rm QC} |  \psi(t) \rangle |^2 $ grows substantially and reaches a maximum as the overlap $|\langle {\rm CDW }| \psi (t) \rangle |^2$ or $|\langle {\bf 0 }| \psi (t) \rangle |^2$ is most suppressed at periodic intervals [Fig.\,\ref{fig:fidelity}\,(c) and (d)], showing there is a periodic transfer of weight from CDW to QC and back. 
The dependence of amplitude on fugacities strongly suggests that the quantum scar state approximates the QC state with small $(\alpha,\beta) \ll 1$, corresponding to a dilute quadrupole density.
\\

{\it Discussion} - The quench dynamics in the dipolar Bose-Hubbard model reveals that correlation spreading is mediated by dipole excitations. Effective models for the dipole dynamics can explain the observed quench dynamics at both integer and half-integer fillings, though in detail they are substantially different in that the Lieb-Robinson velocity is set by $J$ for the integer quench and by $J^2/U$ in the half-integer case. The CDW ground state at half-integer filling bring dramatic changes in the low-energy dipole dynamics. Furthermore, quantum scar states exist in the form of quadrupole excitations in the CDW quench, manifesting themselves as periodic oscillations in the dipole correlation function. The scar-like features in the CDW quench are captured by an effective model resembling the PXP Hamiltonian. The existence of CDW ground state as well as the novel quench dynamics at half-integer filling can be probed in future tilted optical lattice setup.

\paragraph*{Acknowledgments.} H.-Y.L. thanks R. Kaneko for useful discussions at the early stage of the project. Y.-T. O. acknowledges support from the National Research Foundation of Korea (NRF) under grants NRF-2022R1I1A1A01065149. J.H.H. was supported by the NRF grant funded by the Korea government (MSIT) (No. 2023R1A2C1002644). H.-Y.L. and Y.-T.O. were supported by the NRF grant funded by MSIT under grants No. 2020R1I1A3074769 and RS-2023-00220471.

\bibliography{ref}

\bs 

\pagebreak
\bs 

\onecolumngrid
\newpage
\begin{center}

\bs 
\bs 
\textbf{\large Supplementary Material for ``Fractonic Quantum Quench in Dipole-constrained Bosons
''}\\[.3cm]
\end{center}

\setcounter{section}{0}
\setcounter{equation}{0}
\setcounter{figure}{0}
\setcounter{table}{0}
\setcounter{page}{1}
\renewcommand{\theequation}{S\arabic{equation}}
\renewcommand{\thefigure}{S\arabic{figure}}

\bs

\section{Mott phase effective theory}

The Mott-to-Mott quench processes can be understood quite well using the effective model constructed deep within the Mott regime $U\gg J$, where a dilute gas of left and right dipoles dominates the low-energy spectrum. The two kinds of one-dipole states introduced in the main text are given by:
\begin{align}
|l_{x}\rangle = | (n+1)_{x} (n-1)_{x+1}  \rangle , ~~ 
|r_{x}\rangle = | (n-1)_{x} (n+1)_{x+1} \rangle  \nonumber 
\end{align}
in the occupation number basis. Undesignated sites have the occupation $n_x = n$. The Mott state $|M\rangle = |\cdots n_x \cdots \rangle$ serves as the vacuum. The dipole-hopping $J$-term in DBHM acting on $|M\rangle$ creates a pair of $l$ and $r$ dipoles: 
\begin{align}
&b_x (b_{x+1}^\dag)^2 b_{x+2} |M\rangle   = n \sqrt{(n+1)(n+2)}|r_{x} l_{x+1}\rangle ,\nn 
& b_x^\dag (b_{x+1})^2 b_{x+2}^\dag  |M\rangle = (n+1) \sqrt{ n (n-1)}|l_{x} r_{x+1}\rangle, \nonumber 
\end{align}
where
\begin{align}
|r_x l_{x+1} \rangle \equiv & | (n-1)_{x} (n+2)_{x+1} (n-1)_{x+2} \rangle \nn
|l_x r_{x+1} \rangle \equiv & | (n+1)_{x} (n-2)_{x+1} (n+1)_{x+2} \rangle 
\nonumber 
\end{align}
with unmarked sites occupied by $n$ bosons. The dipole pair then drifts apart by further action of dipole hopping. 

In terms of the dipole operators, the dipolar hopping operators can be replaced by
\begin{align}
b_x (b_{x+1}^\dag)^2 b_{x+2} & \rightarrow  n(n+1) \left( l_{x} l_{x+1} ^\dag + r_{x}^\dag r_{x+1}\right)
+ n \sqrt{(n+1)(n+2)} r^\dag_{x} l^\dag_{x+1} 
+ (n+1) \sqrt{n(n-1)} l_{x} r_{x+1},   \nn
b_x^\dag (b_{x+1})^2 b_{x+2}^\dag & \rightarrow  n(n+1) \left( l_{x}^\dag l_{x+1} + r_{x} r_{x+1} ^\dag\right)
+ n \sqrt{(n+1)(n+2)} r_{x} l_{x+1} 
+ (n+1) \sqrt{n (n-1)} l^\dag_{x} r^\dag_{x+1}.
\label{eq:hop-e}
\end{align}
The Hubbard interaction in the dipole subspace becomes
\begin{align}
H_U \equiv U \sum_x (l_x^\dag l_x + r^\dag_x r_x ). 
\label{eq:Hubbard-e}
\end{align} 
This assumes that the dipoles are far apart, and each dipole costs energy $+U$. The dipole creation/annihilation processes take place only when they are adjacent, as indicated by the pair-creation and annihilation terms in Eq.\,(\ref{eq:hop-e}). This, however, is a rare event in the case of dilute-diplon regime, and for the most part the Hubbard energy is simply given by Eq.\,(\ref{eq:Hubbard-e}). In the same dilute-dipole regime, $r$ and $l$ operators can be treated as ordinary boson operators subject to the hard-core constraints $(r^\dag_x )^2 = (l^\dag_x )^2 = 0$. The constraints are, in turn, resolved by mapping the boson model to the fermion model through Jordan-Wigner transformation\,\cite{bloch12,kollath12}. We follow the same footsteps and arrive at the effective Hamiltonian.

In the momentum space, the effective Hamiltonian becomes 
\begin{align}
H_{\rm eff} = & \sum_k [\rho_k ( l_{k}^\dag l_{k} + r_{\overline{k}}^\dag r_{ \overline{k}} ) - \lambda_k ( e^{-i \mu_k } l_{k}^\dag r_{\overline{k}}^\dag - e^{i \mu_k} l_{k} r_{ \overline{k}} ) ], 
\label{eq:dipole-hamil}
\end{align}
where 
\begin{align}
\rho_k & = U - 2 J n (n+1) \cos k, \nn 
\lambda_k e^{i \mu_k } & = J ( n \sqrt{(n\!+\!1)(n\!+\!2)} e^{ik} \!-\! (n\!+\!1) \sqrt{n (n\!-\!1)} e^{-ik} ) . \nonumber 
\end{align}
After the Bogoliubov transformation,
\begin{align}
\gamma_{l,k}^\dag = u_k l_{k}^\dag + v_k r_{\overline{k}}, ~~
\gamma_{r,\overline{k}}^\dag 
=  - v_k l_{k}   + u_k r_{\overline{k}}^\dag, 
\label{eq:bogo-tranf}
\end{align}
where 
\begin{align} u_k =  \cos \theta_k, ~ v_k = \sin \theta_k e^{i\mu_k}, ~ \theta_k = \frac{1}{2}\tan^{-1}\left(- \frac{\lambda_k }{\rho_k }\right), \nonumber \end{align}  
one obtains 
\begin{align}
    H_{\rm eff} = \sum_{k,\sigma} \omega_k \gamma_{k\sigma}^\dagger \gamma_{k\sigma}  \label{eq:dipole-hamil-omega-sm}
\end{align}
with $\omega_k = (\rho_k^2  + \lambda_k^2)^{1/2}$ describing quasiparticle dynamics deep in the Mott phase of DBHM. 

The ground state of the post-quench Hamiltonian is given by $\gamma_{l , k} |M' \rangle = 0$ and $\gamma_{r , k} |M' \rangle = 0$ in the quasiparticle picture, related to the pre-quench ground state (Mott state) $|M\rangle$ by 
\begin{align}
& |M\rangle   = \prod_k \left[ \cos \theta_k + \sin \theta_k e^{-i \mu_k}\gamma_{l,k}^\dag \gamma_{r,\overline{k}}^\dag\right] |M'\rangle.
\end{align}
One can show that $l_k |M\rangle = r_k |M\rangle = 0$.  The time evolution of the post-quench state follows as 
\begin{align}
|\psi(t)\rangle = \prod_k \bigl( \cos \theta_k + \sin \theta_k   e^{-i ( 2\omega_k  t + \mu_k ) } \gamma^\dag_{l,k} \gamma^\dag_{r,\overline{k}} \bigr) |M' \rangle  . 
\label{eq:mott-quenched-state}
\end{align}

\section{Dipole correlator in the Mott phase}

In the TDVP calculation, we calculate the dipole correlator:
\begin{align}
C_d(x, t) = \langle \psi(t) | d_{x_0+x}^\dag d_{x_0} |\psi(t)\rangle,
\end{align}
where $d_x = b_x^\dag b_{x+1}$. We assume that the quenched state $|\psi(t)\rangle$ is represented according to Eq. \eqref{eq:mott-quenched-state}.
By expressing the operator $d_{x_0 + x}^\dag d_{x_0}$ with the dipole operators $l_x$ and $r_x$, the correlator is expressed as
\begin{align}
C_d(x, t) = n (n+1) \left[C_{l^\dag l}(x, t) + C_{r^\dag r}(x, t) + C_{l^\dag r^\dag}(x, t) +C_{l r}(x, t)\right],
\end{align}
where the four dipole correlators are given by
\begin{align}
C_{l^\dag l}(x, t) \equiv & \langle \psi(t) |  l_{x_0+x}^\dag l_{x_0} |\psi(t) \rangle,\nn
C_{r^\dag r}(x, t) \equiv & \langle \psi(t) |  r_{x_0}^\dag r_{x_0+x} |\psi(t) \rangle, \nn 
C_{l^\dag r^\dag}(x, t) \equiv & \, {\rm sgn}(x) \langle \psi(t)| l_{x_0+x}^\dag r_{x_0}^\dag |\psi(t)\rangle, \nn 
C_{l r} (x, t) \equiv & \, {\rm sgn}(x) \langle \psi(t)| l_{x_0} r_{x_0+x}|\psi(t)\rangle,
\end{align}
where ${\rm sgn}(x)$ denotes the sign of $x$.

Performing the Fourier transformation and the Bogoliubov transformation in sequence, one can get
\begin{align}
C_{l^\dag l}(x, t) = & \frac{1}{2 L}\sum_k e^{- i k x} \sin^2 \left(2 \theta_k\right) \left[1 - \cos\left( 2 \omega_k t\right) \right], 
\nn 
C_{r^\dag r}(x, t) =& \frac{1}{2 L}\sum_k e^{- i k x} \sin^2 \left(2 \theta_k\right) \left[1 - \cos\left( 2 \omega_k t\right) \right], 
\nn
C_{l^\dag r^\dag}(x, t) = & -  \, {\rm sgn}(x)\frac{i}{L} \sum_k e^{-i k x - i \mu_k} \sin(2 \theta_k) \sin(\omega_k t) 
\left( e^{i\omega_k t} \cos^2 \theta_k + e^{-i \omega_k t} \sin^2 \theta_k\right), 
\nn
C_{l r}(x, t) = & -  \, {\rm sgn}(x) \frac{i}{L} \sum_k e^{-i k x + i \mu_k} \sin(2 \theta_k) \sin(\omega_k t)
\left( e^{-i\omega_k t} \cos^2 \theta_k + e^{i \omega_k t} \sin^2 \theta_k\right).
\end{align}
Here, $\theta_k$ and $\mu_k$ are parameters for Bogoliubov transformation given in Eq.\,(\ref{eq:bogo-tranf}), and $\omega_k$ is the spectrum of post-quenched Hamiltonian given in Eq.\,(\ref{eq:dipole-hamil-omega-sm}). 

Summarizing all together, the dipole correlator can be expressed as
\begin{align}
C_d(x, t) = \frac{n(n+1)}{L} \sum_k \frac{\lambda_k e^{-i k x}}{\omega_k^2}
\Bigg[&  \lambda_k \left(1 - \cos\left( 2 \omega_k t\right) \right) 
+ 2 i  \, {\rm sgn}(x) \sin\left(\omega_k t\right) 
\left( \omega_k \cos\left( \omega_k t\right) \cos \mu_k + \rho_k \sin\left(\omega_k t\right) \sin \mu_k \right)
\Bigg].
\end{align}

\section{Determination of the group and phase velocities}

\begin{figure*}[h!]
    \centering
    \includegraphics[width=0.95\textwidth]{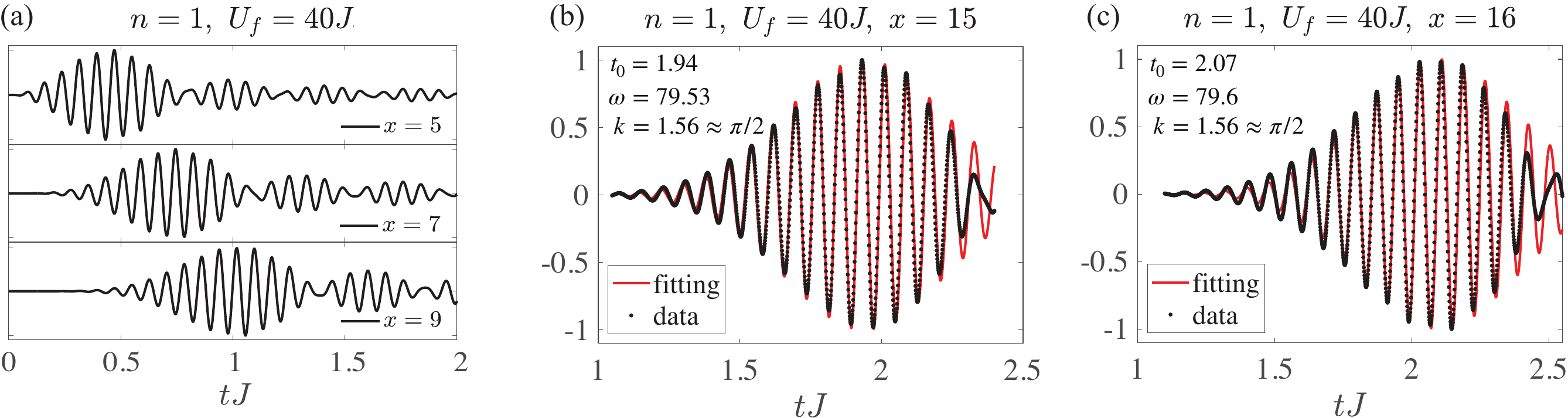}
    \caption{(a) Dynamics of dipole correlation as a function of time at $x=5,7$ and $9$. The leading wavepacket with its fitting function at $x=15$ and $16$ with fitting parameters $t_0$, $\omega$ and $k$.}
    \label{fig:velocity_fitting}
\end{figure*}

Figure\,\ref{fig:velocity_fitting},(a) illustrates the propagation of the dipole correlation, denoted as $C_d(x,t) = {\rm Re}[ \langle \psi(t) | d_{L/2}^\dagger d_{L/2+x} |\psi(t)\rangle ]$, as a function of time at distances $x = 5, 7$, and $9$. We fit the wavepacket appearing earliest to a Gaussian wavepacket,
\begin{align}
    C(x,t) = e^{ - \left( \frac{t-t_0}{\sigma} \right)^2 } \sin( \omega t + k x )   
\end{align}
where the parameters $t_0$, $\omega$, $k$, and $\sigma$ represent the center, frequency, wavenumber, and width of the wavepacket, respectively, which can be determined through fitting at given $x$. We consider $x$ values ranging from 15 to 20, which are sufficiently far from the center yet not close to the boundary. 
For instance, Figure\,\ref{fig:velocity_fitting}\,(b) and (c) display the leading wavepacket along with the fitting function at $x=15$ and $16$, respectively. From $t_0$ values obtained for various $x$, one can estimate the group velocity using the formula: $v_g = (x_2 - x_1) / (t_0(x_2) - t_0(x_1))$. For the phase velocity, we utilize the formula $v_p = \omega / k$, where both $\omega$ and $k$ are obtained from the fitting procedure. 

\section{Filling Number Dependency of Group Velocity at large $U$ limit}
The group velocity from the effective theory is given in Eq.\,(\ref{eq:velocities}) of the main text. More specific form is given by
\begin{align}
v_g  = &  \max_{k = k_{\rm max}} 2 \omega_k 
= \max_{k = k_{\rm max}} \frac{2}{\omega_k}\left( \rho_k \partial_k \rho_k + \lambda_k \partial_k \lambda_k \right) \nn 
= & \max_{k = k_{\rm max}} \frac{4 J n (n+1) \sin k \left(U - 2 J n(n+1) \cos k \right) + 8 J n (n+1)\sqrt{(n-1)n(n+1)(n+2)} \sin 2 k}
{\left( \left(U - 2 J n(n+1) \cos k\right))^2 +
J^2 \left(2n^4 + 4n^3 + n^2 - n - 2 n (n+1)\sqrt{(n-1)n(n+1)(n+2)} \cos 2 k \right)
\right)^{1/2}}.
\end{align}
In the limit of  $U\gg J n^2$, one can deduce $k_{\rm max} \approx\pi /2 $. At $k = k_{\rm max}$, the group velocity is given by
\begin{align}
v_g = &  \frac{4 J n(n+1) U^2}
{\left(U^2 + J^2\left(2n^4 + 4n^3 + n^2 - n + 2 n (n+1)\sqrt{(n-1)n(n+1)(n+2)}  \right)  \right)^{1/2}} \nn 
\approx & 4 J n(n+1) \left( 1 - \frac{J^2}{2 U^2}\left(2n^4 + 4n^3 + n^2 - n + 2 n (n+1)\sqrt{(n-1)n(n+1)(n+2)}  \right)  \right).
\end{align}

\section{Group and Phase Velocities at Fixed $U$ }

\begin{figure}[h!]
    \centering
    \includegraphics[width=0.55\textwidth]{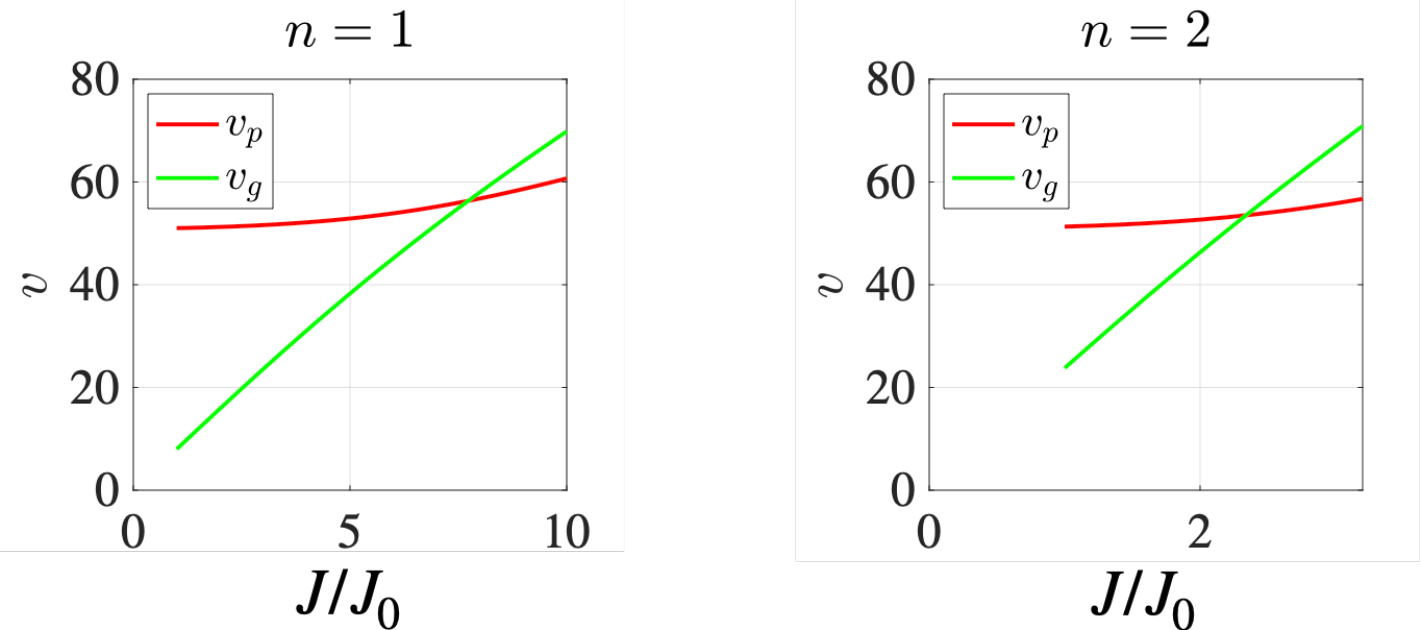}
    \caption{Phase and group velocities for the dipole correlation at $n=1$ and $2$ as a function of $J/J_0$ with constant $U/J_0 = 40$.
    The green and red lines represent the calculated group ($v_g$) and phase ($v_p$) velocities derived from the effective model using Eq.\,(\ref{eq:velocities}) of the main text. }
    \label{fig:sm_velocities}
\end{figure}

Here, we depict the group velocity ($v_g$) and the phase velocity ($v_p$) as a function of $J$ while keeping $U$ fixed. 
In other words, we consider a quench process: $(J_0, U) \rightarrow (J, U)$.
Figure\,\ref{fig:sm_velocities} illustrates the velocities calculated from Eq.\,(\ref{eq:velocities}) of the main text for both $n=1$ and $n=2$ as a function of $J/J_0$, with $U/J_0= 40$ kept constant. Here, $J_0$ serves as the normalized energy scale.
In both cases, one can check the group velocity increases linearly with $J$, whereas the phase velocity remains relatively unaffected by the variations in $J$.

\section{Additional Data on Quench Dynamics in the Mott Phase for $n=1$ and $n=2$ Fillings }

\begin{figure*}[h!]
    \centering
    \includegraphics[width=0.95\textwidth]{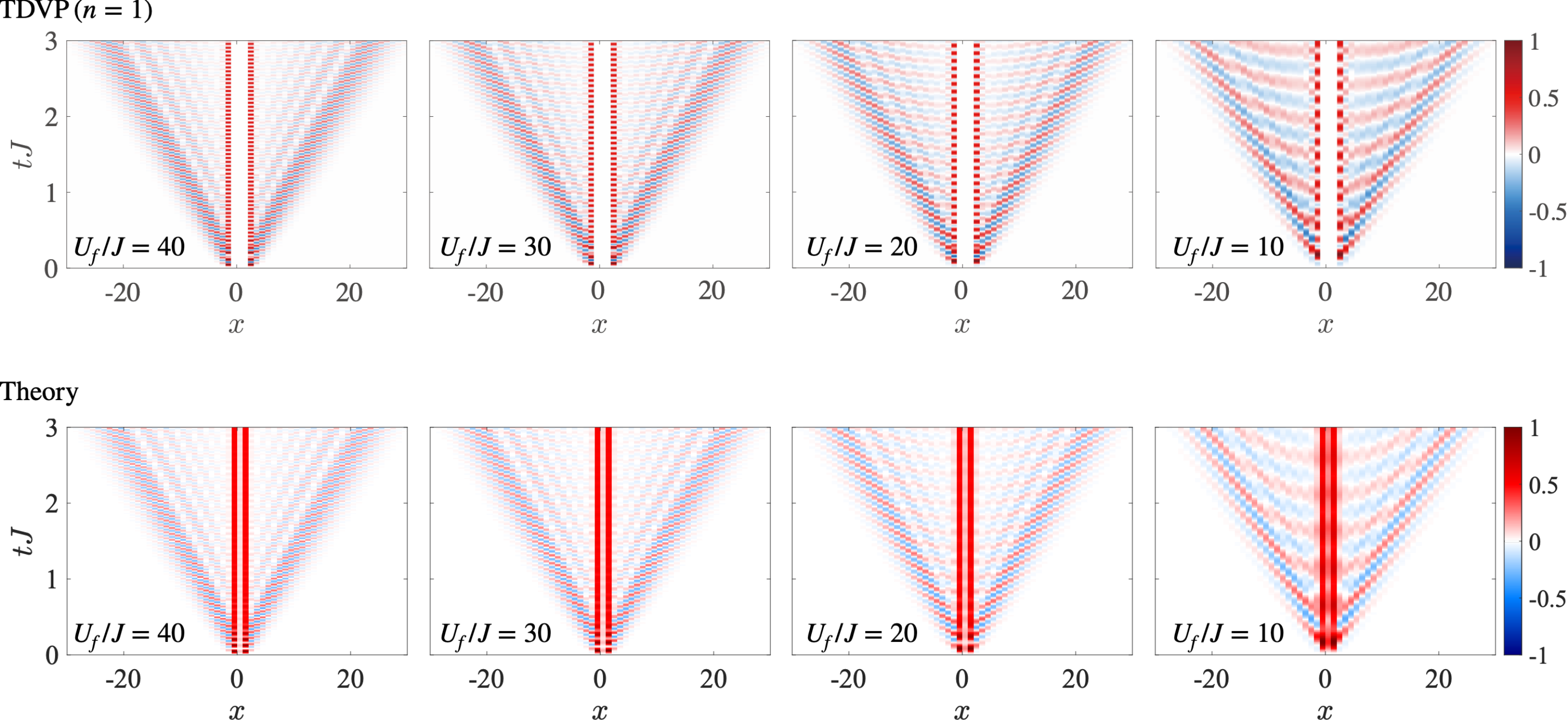}
    \caption{ Spreading of the dipole correlations at the filling $n=1$. Here, the initial state is the ground state at $U_i/J=100$. The data is normalized such that the maximum  value is adjusted to unity.  }
    \label{fig:dipole_correlation_n1}
\end{figure*}

\begin{figure*}[h!]
    \centering
    \includegraphics[width=0.95\textwidth]{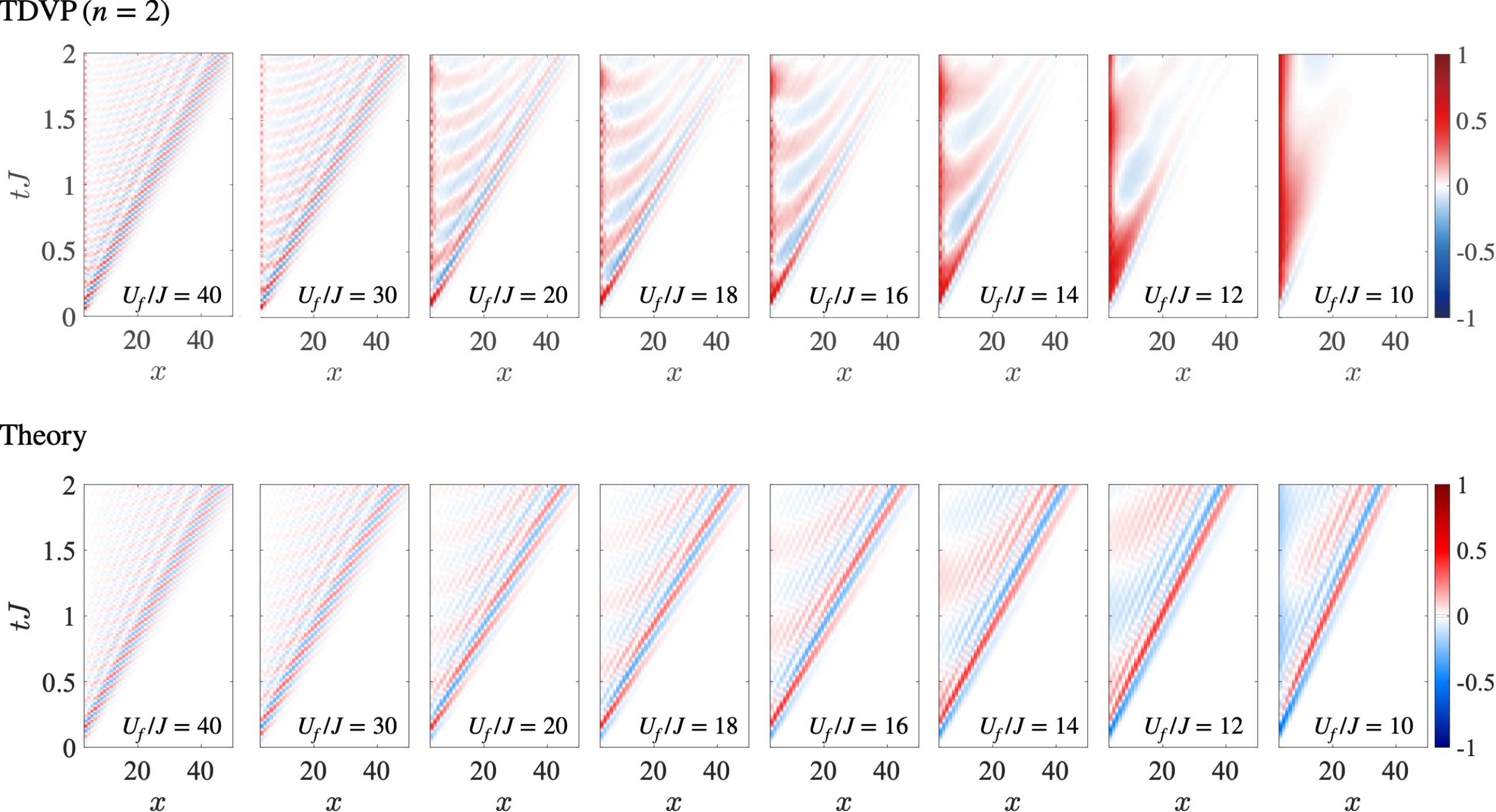}
    \caption{Spreading of the dipole correlations at the filling $n=2$. Here, the initial state is the ground state at $U_i/J=100$. The data is normalized such that the maximum  value is adjusted to unity. }
    \label{fig:dipole_correlation_n2}
\end{figure*}

Here, we provide supplementary results from the TDVP simulations of quench dynamics within the Mott phase of the DBHM for two specific filling factors: $n=1$ and $n=2$. Additionally, we compare these results with the ones obtained from the effective theory. See Figs.\,\ref{fig:dipole_correlation_n1} and \ref{fig:dipole_correlation_n2}.


\begin{figure*}[h!]
    \centering
    \includegraphics[width=0.6\textwidth]{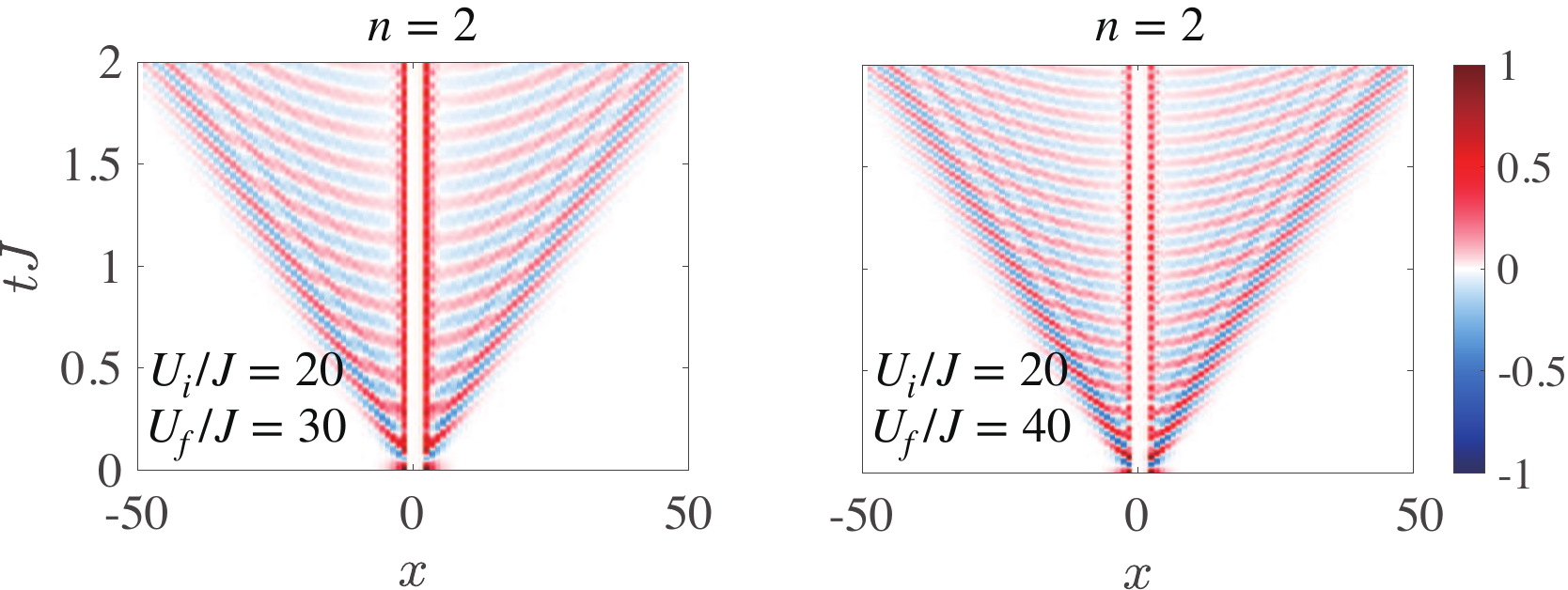}
    \caption{Propagation of the dipole correlations at the filling $n=2$. Here, the quench direction is reverse, i.e., from a smaller $U_i/J=20$ to larger (left) $U_f/J=30$ and (right) $U_f/J=40$. The data is normalized such that the maximum  value is adjusted to unity.}
    \label{fig:mott_quench_opposite}
\end{figure*}

Additionally, we include TDVP data for the quench dynamics transitioning from a smaller initial on-site interaction strength $U_i$ to a larger quench strength $U_f$ within the Mott phase, as illustrated in Fig.\,\ref{fig:mott_quench_opposite}. Our findings indicate a propagation speed of the dipole correlation is consistent with that from the reverse quench direction discussed in the main text, where $v_g$ is approximately 20. This observation corroborates the effective theory's prediction that the group velocity depends only on the hopping strength $J$.

\section{CDW phase perturbation theory}

The ground state space of $H_0$ at half-integer filling ($\nu = 3/2$) is denoted ${\cal W}$. Having identified the CDW state as reference vacuum state in ${\cal W}$, other low-lying excited states in ${\cal W}$ can be identified with the creation of $l$ and $r$ dipoles above the CDW. A subset of states in ${\cal W}$ is derived by transferring two bosons from two next-nearest neighboring sites within a CDW state that each hosts $2$ bosons—one moving to the left and the other to the right:
\begin{align}
    |\cdots , {\color{magenta} 1 , 2} , 1 , {\color{cyan} 2, 1}, \cdots \rangle 
    & \rightarrow |\cdots , {\color{magenta} 2, 1}, 1, {\color{cyan} 1, 2}, \cdots \rangle , \nn 
    |\cdots , {\color{cyan} 2 , 1} , 2 , {\color{magenta} 1, 2}, \cdots \rangle 
    & \rightarrow |\cdots , {\color{cyan} 1, 2}, 2 , {\color{magenta} 2, 1} , \cdots \rangle  . 
\end{align}
In each scenario, a pair of $1,2$ (magenta) and $2,1$ (cyan) swaps to form $2,1$ and $1,2$ pairs, respectively. The transitioned pair to $2,1$ is designated as an $l$ dipole, and the $1,2$ pair as an $r$ dipole, each seamlessly embedded into the otherwise perfect CDW pattern. A notable distinction from the Mott phase with integer filling is that the $l$ dipole is exclusively generated and located on odd-numbered sites, whereas the $r$ dipole is restricted to even-numbered sites. Considering the dipole-moment conservation, the numbers of $l$ and $r$ dipoles equal in the basis of ${\cal W}$.

The ground state space of $H_0$ at half-integer filling ($\nu = 3/2$) is denoted ${\cal W}$, and a particular state in ${\cal W}$ is $|\psi^{\cal W}_\alpha \rangle$. For any two states within this subspace we have $\langle \psi_\alpha^{\cal W} | H_1  |\psi_\beta^{\cal W} \rangle = 0$, since $H_1$ changes the occupation at a particulate site by two (e.g. $2 \rightarrow 0$ or $1 \rightarrow 3$) and lifts the state out of ${\cal W}$. We thus need to employ $H_1$ to second order to lift the degeneracy.

The effective Hamiltonian for subspace ${\cal W}$, as derived using second-order perturbation theory, is given by
\begin{align}
H^{\cal W}_{\rm eff}  = \sum_{\alpha,\beta \in {\cal W}} |\psi_\alpha^{\cal W} \rangle h_{\alpha \beta}^{(2)} \langle \psi_\beta^{\cal W}|
\label{eq:cdw-2nd-h-1}
\end{align}
where
\begin{align}
h^{(2)}_{\alpha\beta} =  \sum_p \frac{1}{E_{\cal W}^{(0)} - E_p^{(0)}}   \langle \psi_\alpha^{\cal W}| H_1 | p^{(0)}\rangle \langle  p^{(0)} | H_1 |\psi_\beta^{\cal W} \rangle.
\label{eq:cdw-2nd-h-2}
\end{align}
Here, $|p^{(0)}\rangle$ are the first excited states of $H_0$ lying outside ${\cal W}$. It can be recognized that the second-order Hamiltonian quantifies how much a state within a given subspace temporarily shifts to a state outside that subspace, due to the effects of $H_1$, before transitioning to a different state within the original subspace.

In the second-order perturbation by $H_1$, we have two pathways that moves $r$ dipole to the right:
\begin{align}
|\cdots,{\color{cyan} 1 , 2}, 2, 1 , 2, \cdots \rangle
\rightarrow &  |\cdots, 2 ,0 , 3, 1 , 2, \cdots  \rangle
\rightarrow    |\cdots, 2, 1, {\color{cyan} 1, 2}, 2, \cdots \rangle,  
\nn 
|\cdots, {\color{cyan} 1 , 2}, 2, 1 , 2, \cdots \rangle
\rightarrow  & |\cdots, 2, 3, 0, 2 , 2, \cdots \rangle
\rightarrow    |\cdots, 2, 1, {\color{cyan} 1, 2}, 2, \cdots\rangle, 
\end{align}
moving the defect position by two sites to the right. The energy gap between the initial (and final) states and the intermediate state amounts to $2U$, leading to $E_{\cal W}^{(0)} - E_p^{(0)} = -2U$. In each pathways, it is verifiable that $\langle \psi_\alpha^{\cal W}| H_1 | p^{(0)}\rangle \langle p^{(0)} | H_1 |\psi_\beta^{\cal W} \rangle = 12 J^2$. Similarly, moving an $l$ dipole results in equivalent outcomes in the second-order Hamiltonian as shown in Eq.\,(\ref{eq:cdw-2nd-h-2}), establishing the transition amplitude for the hopping of either $l$ or $r$ dipoles as
\begin{align}
h^{(2)}_{l_{2a-1} r_{2b}, l_{2a-3},r_{2b}}=
h^{(2)}_{l_{2a-1} r_{2b}, l_{2a+1},r_{2b}} = h^{(2)}_{l_{2a-1} r_{2b}, l_{2a-1},r_{{2b}\pm 2}} = - 12 \frac{J^2}{U}. 
\end{align}

Subtracting the second-order energy shift of the CDW state from this value gives the effective on-site energy of the $l$ and $r$ dipoles. In calculating the on-site terms, $h^{(2)}_{\alpha\alpha}$, we now have to consider eight pathways for these on-site terms. Omitting the detailed derivation, the on-site energy for a single pair of $l$-$r$ dipoles is calculated as
\begin{align}
h^{(2)}_{l_{2a-1} r_{2b}, l_{2a-1} r_{2b} } - h^{(2)}_{{\rm CDW}, {\rm CDW}} = \frac{328}{5} \frac{J^2}{U} = 65.6 \frac{J^2}{U} > 0.
\end{align}

It's observed that the creation and annihilation of $l$-$r$ dipole pairs don't manifest even in the second-order perturbation. However, it's possible to construct an effective Hamiltonian that accounts for $l$-$r$ pairs being generated at the second-order level. Combining all the considerations, we arrive at the effective model for the dipole excitations: 
\begin{align}
H^{D}_{\rm eff} = & - 12\frac{ J^2}{U} \sum_{a} \left(\, |l_{2a-1}\rangle \langle l_{2a+1}| + |r_{2a}\rangle \langle r_{2a+2}| + h.c. \right) \nn 
    &+32.8\frac{J^2}{U} \sum_a \left(|l_{2a-1}\rangle \langle l_{2a-1}| + |r_{2a} \rangle \langle r_{2a} | \right) .
\label{eq:cdw-lr}
\end{align}
\section{Quadrupole condensate states}

We employ the matrix product state representation\,(MPS) to construct a quadrupole condensate\,(QC) state:
\begin{align}
    |{\rm QC}\rangle 
    = & \Omega \, P \, \prod_{x \in {\rm odd}} \left( 1 + \alpha \, X_x \right) \prod_{x \in {\rm even}} \left( 1 + \beta \, X_x \right)|{\bf 0}\rangle \quad \quad \quad \quad \quad \quad \quad {\rm (in\,\,PXPQ)} \nn 
    = &  \Omega \, P \, \prod_{x\in {\rm odd}} \left( 1 + \frac{\alpha}{2\sqrt{6}}\,\, q_x^\dagger \right) \prod_{x\in {\rm even}} \left( 1 + \frac{\beta}{2\sqrt{2}}\,\, q_x^\dagger \right) |{\rm CDW}\rangle \quad {\rm (in\,\, DBHM)} 
\end{align}
where a projector $P$ rules out configurations that contain neighboring two 202 excitations or neighboring 131-202 excitations. Here, $\alpha$\,($\beta$) is the fugacity of the 131\,(202) excitation.  Taking the 2-site unit-cell structure, the MPS consists of two different tensors, $T_a$ and $T_b$ such that the wavefunction of the QC state is given as
\begin{align}
    \Psi_{\rm QC}^{n_1 n_2 \cdots n_L} = {\rm tTr}\left[ T_a^{n_1} T_b^{n_2} T_a^{n_3} \cdots T_b^{n_L} \right],
\end{align}
where ${\rm tTr}[\cdots]$ denotes the tensor trace, and $n_x$ stands for the local boson number, and $L$ is the system size being an even number. There are two types of local quadrupole, i.e., $|131\rangle_{x}$\,(or $|1\rangle_x$) and $|202\rangle_{x}$\,(or $|1\rangle_x$) at even and odd sites, respectively. In DBHM, the QC state for general $(\alpha, \beta)$ can be generated by two tensors $T_a$ and $T_b$ with the bond dimension $\chi=4$ defined as below:
\begin{align}
    & [T_a]_{0,0}^1 = [T_a]_{0,3}^2 = [T_a]_{3,0}^2 = 1, \quad
    [T_a]_{1,1}^3 = [T_a]_{2,2}^3 = \alpha, \nonumber \\
    & [T_b]_{0,0}^2 = [T_b]_{0,1}^1 = [T_b]_{1,0}^1 = [T_b]_{2,0}^1 = [T_b]_{1,2}^0 = 1, \quad 
    [T_b]_{3,3}^0 = \beta.
\end{align}
On the other hand, in the PXPQ model, the QC state can be generated by two tensors $T_a$ and $T_b$ with the bond dimension $\chi=3$ defined as below:
\begin{align}
    & [T_a]_{0,0}^0 = [T_a]_{1,1}^0 = [T_a]_{2,2}^0 = 1, \quad
    [T_a]_{0,0}^1 = \alpha, \nonumber \\
    & [T_b]_{0,0}^0 = [T_b]_{0,1}^0 = [T_b]_{2,0}^0 = 1, \quad 
    [T_b]_{1,2}^1 = \beta.
\end{align}

\end{document}